\documentstyle[aps,prb,jpc,preprint]{revtex}
\tightenlines
\begin{document}

\title{Molecular theories and simulation of ions and polar molecules
  in water}

\author{Gerhard Hummer, Lawrence R. Pratt, Angel E. Garc\'{\i}a }

\address{Theoretical Division, Los Alamos National Laboratory, Los
Alamos, New Mexico 87545 USA }

\author{LA-UR-98-1947}

\date{\today}

\maketitle

\clearpage
\begin{abstract}
Recent developments in molecular theories and simulation of ions and
polar molecules in water are reviewed.  The hydration of imidazole and
imidazolium solutes is used to exemplify the theoretical issues.  The
treatment of long-ranged electrostatic interactions in simulations is
discussed extensively.  It is argued that the Ewald approach is an
easy way to get correct hydration free energies in the thermodynamic
limit from molecular calculations; and that molecular simulations with
Ewald interactions and periodic boundary conditions can also be more
efficient than many common alternatives.  The Ewald treatment permits
a conclusive extrapolation to infinite system size.  Accurate results
for well-defined models have permitted careful testing of simple
theories of electrostatic hydration free energies, such as dielectric
continuum models.  The picture that emerges from such testing is that
the most prominent failings of the simplest theories are associated
with solvent proton conformations that lead to non-gaussian
fluctuations of electrostatic potentials.  Thus, the most favorable
cases for second-order perturbation theories are monoatomic positive
ions.  For polar and anionic solutes, continuum or gaussian theories
are less accurate.  The appreciation of the specific deficiencies of
those simple models have led to new concepts, multistate gaussian and
quasi-chemical theories, that address the cases for which the simpler
theories fail.  It is argued that, relative to direct dielectric
continuum treatments, the quasi-chemical theories provide a better
theoretical organization for the computational study of the electronic
structure of solution species.

\end{abstract}
\pagebreak

\section{Introduction}
Water, the most commonly encountered liquid, exerts both chemical and
physical influences on aqueous molecular processes.  Hydration effects
are often divided into hydrophobic and hydrophilic categories.
Hydrophilic solutes are typically ionic or polar species and may
participate in chemical interactions with the water solvent.  Because
of the long range of the electrostatic interactions and their strength
relative to $k_{\rm B}T$, hydrophilic hydration presents distinctive
conceptual and practical issues for understanding and predicting the
influence of hydration on chemical and biochemical events in water.

A principal and long-standing technical issue is the treatment of
infinitely long-ranged interactions in the context of a sample of
finite size.\cite{stell} Recent work has helped to resolve this
problem.  One algorithmic approach to treatment of long-ranged
interactions is the use of Ewald interactions within the conventional
periodic boundary conditions.\cite{Ewald:21} We argue here that {\em
the Ewald approach is an easy way to get correct hydration free
energies} from molecular calculations, that is, to achieve well
characterized results appropriate to the thermodynamic limit in which
the system size tends to infinity for given densities and temperature.
What is more: molecular simulations with Ewald interactions and
periodic boundary conditions can also be more efficient than rougher
approximations that are often employed to compute hydration free
energies for molecularly well-defined problems.  We anticipate results
below by noting that we obtain accurate, thermodynamic limiting
results for the hydration free energy of imidazole with as few as 16
water molecules included in the simulation.  The price to be paid for
this accuracy and efficiency is additional effort in understanding
Ewald calculations from a physical viewpoint and in implementing Ewald
interactions,\cite{Ewald:21} its
equivalents,\cite{Slattery:80,Lekner:91,Darden:93,Schmidt:91,pollock}
and alternatives.\cite{Barker:73,Hummer:92,Hummer:94:e}

The physical issues motivating simulation calculations of this type
revolve around dielectric continuum models of hydration of ionic and
polar solutes.\cite{Gilson:95} It is natural and common for a
simplified approximation to provide a conceptual baseline for
considering more accurate theoretical results.  But the converse
comparison is foremost for this work.  The theoretical efforts over
recent years have provided sharper tests of the validity of the
continuum approach than merely: is an empirically correct hydration
free energy obtained?  Recent work has clarified that the dielectric
models are simple implementations of thermodynamic perturbation theory
through second-order in electrostatic coupling parameters such as
solute charges;\cite{Pratt:94:a,Hummer:95:e,Hummer:96:a} dielectric
models can also be considered a simple implementation of an {\em
ansatz} that electrostatic potential fluctuations are distributed
according to a gaussian probability density,\cite{Levy:91} or they can
be considered a simplified linear response
theory.\cite{Hwang:87,Jayaram:89,Aqvist:96}

Second-order perturbation theory was found to be satisfactory for some
solutes such as alkali ions,\cite{Hummer:96:a} but unsatisfactory for
water\cite{Hummer:95:e,Rick:94} and anions.\cite{Hummer:96:a} In the
latter cases, of course, an {\em a posteriori} adjustment of cavity
radii could still produce the correct hydration free
energies.\cite{Tawa:95} However, the more ambitious molecular theory
ties the values of radii parameters to molecular properties that
depend on the thermodynamic state of the system (temperature,
pressure, and composition of the solvent) and to non-electrostatic
characteristics of the solute-solvent interactions.  The radii are not
separately adjustable when viewed from that deeper level of molecular
theory.
However, the radii can be well-defined and are {\em not}
properties of the solutes alone but incorporate information about the
solvent and thermodynamic state.

For water as a solvent, the case of exclusive concern here, the most
prominent failings of second-order perturbation theory are associated
with solvent proton conformations that lead to non-gaussian
fluctuations of electrostatic potentials.\cite{Hummer:96:a,Hummer:97c}
Thus, the most favorable cases for the second-order perturbation
theories are classic positive ions.  In such cases, oxygen-hydrogen
bonds of water are oriented away from the ion.  Neutral, polar
molecules that may form specific hydrogen bonds with the solvent are
more challenging for these theories, though the hydration free
energies sought are smaller in magnitude than for typical ions.
Negative molecular ions are expected to offer further complications
because now the problematic proton interactions with the solute will
be strong.  However, we have less experience with realistic negative
ions partly because the molecular models used for simulation are less
well developed than for other cases.

The appreciation of the different possibilities for fluctuations has
led to new theories of electrostatic hydration free
energies.\cite{Hummer:97c} These theories analyze electrostatic
distributions more broadly, still using gaussian models at crucial
steps;  but now several gaussian distributions are derived
from an analysis of the first shell environment of the solute.
For the important case of hydration of a water molecule, this
extension repairs the breakdown of a single gaussian
theory.  Negative ions can still be problematic but the multiple
gaussian approach has also motivated development of quasi-chemical
theories\cite{pratt:98} that are based, in principle, on full
information about the thermal motion of the first hydration shell.
Though experience with the quasi-chemical theories is
limited,\cite{Martin:97} we anticipate that they should provide better
descriptions of the hydration free energies, in addition to providing
a reasonable pathway to carry-out solution phase electronic structure
calculations on hydrated negative ions, calculations that would be
difficult particularly in the absence of hydration effects.

In the following section, we will first introduce the model solute
imidazole, which was chosen as a molecular solute to exemplify,
combine, and extend aspects of ionic and polar solutes studied
previously.\cite{Pratt:94:a,Hummer:95:e,Hummer:96:a,Hummer:97c,Garde:98}
Results for imidazole and imidazolium will be used throughout the
manuscript to illustrate the theoretical issues.  We will discuss
the Ewald treatment of electrostatic interactions, motivating it in
various ways.  Subsequently, finite-size effects will be studied.  The
correction for the typically large finite-size effects is essential
for accurate calculations of solvation free energies of polar and
charged solutes.  We distinguish between electrostatic finite-size
effects that are independent of the thermodynamic state and the
characteristics of the solute, and the remaining thermodynamic
finite-size effects.  We will then introduce perturbative methods for
calculating solvation free energies that are based on the
approximately gaussian character of the electrostatic potential
fluctuations.  Non-gaussian behavior and its accurate treatment using
multistate gaussian and quasi-chemical models will be the focus of the
last section.

\section{Example: Imidazole and Imidazolium in Water}
To illustrate the various issues arising in calculations of solvation
free energies of charged and polar molecules, we present new
calculations of the hydration of imidazole and imidazolium. We choose
this example because recent interest in these problems has focused on
predicting acid-base equilibria of biochemical
relevance.\cite{Warshel:86,Bashford:90,Merz:91,DelBuono:94,Saito:95,%
You:95,Figueirido:96,Warshel:EnzymeBOOK} We will calculate the
charging free energies of the protonated imidazolium and the neutral,
polar imidazole (Figure \ref{fig:imidazole-scheme}).  Imidazole in
water provides a rich example: the polar imidazole molecule can be
protonated at the N$_3$ position to form a molecular cation,
imidazolium.  This protonation reaction has a p$K_a$ of about
7.\cite{Ganellin:Jerusalem:74} It provides a basis for p$K_a$
calculations of ionizable residues of
proteins.
\cite{Warshel:86,Bashford:90,Merz:91,DelBuono:94,Saito:95,You:95,%
Figueirido:96} Imidazole is the building block of histidine, one of
the most active amino acids enzymatically and ubiquitous in the active
sites of enzymes that operate at room temperature and neutral
pH.\cite{Warshel:EnzymeBOOK}

The protonation of imidazole has been studied previously using
combinations of dielectric, quantum mechanical, and computer
simulation
methods.\cite{Rashin:JACS:90,Nagy:93,topol,Bash:PNAS:96,Ho:JPC:96}
Here, we will focus on the solvation contribution to the protonation
equilibrium and reserve quantum mechanical intramolecular effects for
subsequent treatment.

The imidazole/imidazolium system is more complex than the systems we
have studied before using explicit solvent models: mono- and divalent
ions,\cite{Hummer:96:a,Hummer:97c} water in
water,\cite{Hummer:95:e} and tetramethylammonium.\cite{Garde:98}
The
analysis below should also illustrate how the calculation of solvation
free energies using the Ewald method and equilibrium fluctuations of
electrostatic potentials can be extended to proteins, in particular
the calculation of p$K_a$'s of amino acids.

{\bf Monte Carlo computer simulations of imidazole in water}. We
studied the solvation of imidazole [Im(p)] and imidazolium [Im(+)] in
water using Monte Carlo (MC) simulations in the canonical ensemble.
For the Im(p) and Im(+) molecules, we used the partial charges and
geometry of Topol et al.,\cite{topol} as compiled in Table
\ref{tab:imidazole}.\nocite{topol,Cornell:JACS:95,Berendsen:87}
For water, we used the SPC/E
model.\cite{Berendsen:87} The temperature was 298 K.  The Ewald method
was used for the long-range electrostatic interactions with a
real-space screening factor of $\eta=5.6/L$, where $L$ is the length
of the periodically replicated cubic box.  A cutoff of $k^2\leq 38
(2\pi/L)^2$ was applied in Fourier space, resulting in $2\times 510$
${\bf k}$ vectors being considered.  A cutoff of $L/2$ was applied to
the Lennard-Jones and real-space electrostatic interactions based on
atoms.  The background dielectric constant in the Ewald method was
corrected from infinity to 80.\cite{deLeeuw:80:a} The partial molar
volume of the imidazole was chosen as two times that of bulk water at
a density of 997.07 kg/m$^3$, such that the
pressure\cite{Hummer:JCP:98} was about one atmosphere.

The Metropolis MC method was used to sample configurational space in
the simulations.\cite{Metropolis:53} The translational and rotational
move widths of water were chosen to give about 40 {\%} acceptance
ratios.  The solute was allowed to move as well.  Simulations started
from random configurations or configurations of previous runs with
different charges, equilibrated for at least 100,000 and 50,000 MC
passes, respectively, where one pass is one attempted move for each of
the particles.  Electrostatic potentials at solute atom positions and
binding energies of the solute were calculated after every pass using
the Ewald method in simulations extending over 200,000 MC passes each.
Simulations were performed for the uncharged, half-charged, and fully
charged Im(p) and Im(+) in their respective geometries.  The
equilibrium simulations were performed with 16, 32, 64, 128, 256, and
512 water molecules to study finite size effects.

To complete the thermodynamic cycle and check for consistency, two
runs of slow growth thermodynamic integration were used to calculate
the free energy of converting the geometry from the uncharged Im(p) to
the uncharged Im(+) conformation within 150,000 MC passes.  Six runs
of 200,000 MC passes were used to calculate the free energy of
converting the polar Im(p) into an Im(+) cation, starting from
different equilibrated configurations and averaging three charging and
three uncharging runs.  The thermodynamic integrations were carried
out with 256 water molecules.

We will discuss the results of these calculations as they come-up in
the theoretical narrative.  However, before considering more subtle
issues we can make a direct comparison of the average electrostatic
potential exerted by the solvent observed during the simulation with
the corresponding predictions of dielectric models.
Figure~\ref{fig:compare}\nocite{Nina,topol,Truong,Pratt:97}
shows that comparison for several sets of
radii in current use.  Such a comparison illustrates the basic issue
of sensitivity of thermodynamic results  to the radii parameters
and whether extant empirically adjusted radii are transferable to
slightly different cases.

\section{Noteworthy Aspects of the Ewald Treatment of Electrostatic
Interactions} Viewed from an historical perspective, the most
appropriate treatment of electrostatic interactions for simulation
calculations has been a contentious issue. In this setting it is
helpful to note some broad, and non-technical, characteristics of
Ewald treatments that might be typically overlooked.  We preface these
observations by noting that simulation calculations treat finite
systems.  Most commonly, periodic (or Born-von Karman) boundary
conditions \cite{ashcroft,peierls} are utilized for the exterior
boundary of the finite system considered.  The theoretical issues
engendered by these boundary conditions with
finite-ranged\cite{pratt:81a,pratt:81b} and long-ranged interactions
are reasonably well understood. Of course, simulation calculations
need not address issues of what is happening outside the simulation
cell.  We note that it is possible to compute the Ewald potential --
and in more than one way -- without consideration of image charges
outside the simulation cell.  So intuitive arguments based upon image
charges can be avoided completely.

It is convenient to express the Ewald electrostatic energy of a system
of partial charges $q_{i\alpha}$ at positions ${\bf r}_{i\alpha}$ on
molecules $i$ as a sum of effective pair interactions and self terms,
\begin{eqnarray}
  \label{eq:Uew}
  U & = & \sum_{i,j \atop i<j}\sum_{\alpha,\beta}
  q_{i\alpha} q_{j\beta}
  \varphi({\bf r}_{i\alpha j\beta})\nonumber\\&& + \sum_{i}
  \sum_{\alpha,\beta \atop \alpha<\beta} q_{i\alpha} q_{i\beta}
  \left[\varphi({\bf r}_{i\alpha
  i\beta})-\frac{1}{|{\bf r}_{i\alpha i\beta}|}\right]\nonumber\\&&
 + \frac{1}{2} \sum_{i}
  \sum_{\alpha} {q_{i\alpha}}^2 \lim_{{\bf r}\rightarrow 0}
  \left[\varphi({\bf r})-\frac{1}{|{\bf r}|}\right]~,
\end{eqnarray}
where ${\bf r}_{i\alpha j\beta}={\bf r}_{j\beta}-{\bf r}_{i\alpha}$.
The Coulomb energy $U$ was split into intermolecular, intramolecular
and self interaction contributions.  
The Fourier representation of $\varphi({\bf r})$ reveals the
periodicity of this potential:\cite{Brush:66}
\begin{eqnarray}
  \label{eq:phi_Four}
  \varphi({\bf r}) = \frac{1}{V}\sum_{{\bf k} \atop k\neq
  0}\frac{4\pi}{{\bf k}^2} e^{i {\bf k}\cdot{\bf r}}~.
\end{eqnarray}
The ${\bf k}$ sum extends over the reciprocal lattice derived from the
real-space lattice ${\bf n}$ of periodically replicated simulation
boxes.  For a cubic lattice of length $L=V^{1/3}$, we have ${\bf
  n}=L\,(i,j,k)$, and ${\bf k}=2\pi L^{-1}\,(i,j,k)$, where $i$, $j$,
and $k$ are integers.  For numerical convenience, $\varphi({\bf r})$
is partly transformed into real space, which leads to its Ewald
lattice sum representation:
\begin{eqnarray}
  \label{eq:phi}
  \varphi({\bf r}) & = & \sum_{\bf n}\frac{{\rm erfc}(\eta|{\bf r} +
    {\bf n}|)} {|{\bf r}+{\bf n}|} + \sum_{{\bf k} \atop k\neq 0}
\frac{4\pi}{V k^2} e^{ -k^2/4\eta^2 + i{\bf k}\cdot{\bf r}}
-\frac{\pi}{V \eta^2}~.
\end{eqnarray}
$\eta$ is a convergence parameter that is chosen to accelerate
numerical convergence.  Note that the value of $\varphi({\bf r})$ is
independent of $\eta$,\cite{Hummer:95:CPL}
\begin{eqnarray}
  \label{eq:dphideta}
  \frac{\partial \varphi({\bf r})}{\partial\eta} \equiv 0~,
\end{eqnarray}
and that the average potential in the box is
zero,\cite{Hummer:96:a,Brush:66,Hummer:93,Nijboer:88}
\begin{eqnarray}
  \label{eq:avepot}
  \tilde{\varphi}({\bf k}=0) & = & \int_V d{\bf r} \varphi({\bf r}) = 0~.
\end{eqnarray}
In the following, we will separate the Coulomb energy $U_{\rm el}$ of
the solute from the total Coulomb energy eq~\ref{eq:Uew}.  When the
system contains a solute with partial charges $q_\beta$ at positions
${\bf r}_\beta$, its electrostatic interaction energy can be split
into a solvent term and self-interactions,
\begin{eqnarray}
  \label{eq:Uew_sol}
  U_{\rm el} & = & \sum_\beta \sum_i \sum_\alpha q_\beta q_{i\alpha}
  \varphi({\bf r}_{\beta i\alpha})\nonumber\\&& +
  \sum_{\beta,\gamma \atop \beta<\gamma} q_\beta q_\gamma
  \left[\varphi({\bf r}_{\beta\gamma})-\frac{1}{|{\bf
  r}_{\beta\gamma}|}\right]\nonumber\\&&
+ \frac{1}{2} \sum_\beta {q_{\beta}}^2 \lim_{{\bf r}\rightarrow 0}
  \left[\varphi({\bf r})-\frac{1}{|{\bf r}|}\right]~.
\end{eqnarray}
The first, second, and third sum are the direct interactions with
water, the interactions of charges on the solute with other solute charges,
and the self-interactions of solute charges, respectively.

{\bf The Ewald potential $\varphi({\bf r})$ is the solution of the
  Poisson equation that is periodic with the fundamental period of the
  simulation cell}.\cite{Hummer:96:a,Brush:66,Hummer:93,Cichocki:89} A
periodic solution of the Poisson equation requires that the surface
integral of the electric field normal to the surface of the simulation
cell be zero.  This means that the material in the simulation cell
must have zero net electric charge. If the physical system of interest
is non-neutral, a uniform background distribution of charge is
included to neutralize the non-zero charge of the physical system.

Consider an elementary charge.  Because the Ewald potential is
periodic, we can consider the Ewald electrostatic potential implied by
centering the simulation cell on this elementary charge.  By symmetry
the Ewald normal electric field is zero on the cell boundary.  The
Ewald potential can thus be considered to be that of a cut-off on the
cell boundary -- the cut-off at the maximum distances achievable --
with zero normal derivative analogous to a shifted-force correction.

{\bf The Ewald potential pushes the electrostatic boundary outward as
  far as possible but still retains smoothness on the boundary}. The
minimum image cut-off shares with the Ewald treatment the property
that the electrostatic potential is not cut-off in any region of the
simulation volume, the largest volume that must be physically
considered.  However, as demonstrated above, the Ewald potential is
smooth on that boundary since it is the periodic solution of the
Poisson equation.  This is an important technical advantage that
facilitates investigation of system size dependence of computed
properties, {\it i.e.}, the variation of system properties with
variations of the cell boundaries.  In those cases where the physical
system of interest is non-neutral it is a helpful point of view that
the background charge density is a simple device that permits
smoothness of the computed potential on the system boundary.  We
emphasize that the effect of the neutralizing background charge
disappears as the thermodynamic limit is approached, in which the
background charge density disappears.

{\bf Reaction field potentials are Ewald potentials for different
background charges.} Reaction field
methods\cite{Barker:73,Hummer:92,Neumann:83:b,Hummer:94:e} are
computationally efficient alternatives to Ewald summation.  The
effective potentials of the site-site reaction field (SSRF)
method\cite{Hummer:92} and the generalized reaction field (GRF)
method\cite{Hummer:94:e} can be viewed as Ewald potentials for a
non-homogeneous background charge.  The SSRF and GRF potentials are
the solutions to the Poisson equation for a source charge and a
compensating background.  The background charge densities in the SSRF
and GRF method are that of a homogeneous sphere and a radially
symmetric charge distribution centered around the source charge,
respectively.  The SSRF and GRF method have a finite range because of
the radially symmetric background charge densities that exactly
compensate the source charge when the cut-off distance is reached.

{\bf Comparison of Ewald potentials from simulation of water to
  electrostatic potentials in isolated water droplets}. It is
interesting to make some simple numerical comparisons between the
Ewald potentials that are experienced in simulation of water with
periodic boundary conditions to the corresponding electrostatic
potentials in water droplets.
Such a comparison is simplified if we locate a distinguished solute at
the origin of our Cartesian coordinate system.  For a spherical
Lennard-Jones solute,
Figure~\ref{fig:cluster}\nocite{Berendsen:81,Hummer:97} shows a
typical variation of the electrostatic potential at the solute center
with inclusion of the charge density in progressively larger spherical
volumes of radius $R$ around the solute.  Notice the substantial
variation of the electrostatic potential with inclusion of the
solvation shells near the solute.  However, after about three shells
the net electrostatic potential oscillates about the Ewald asymptotic
value before the ball penetrates the physical interface of the
droplet.  Thereafter, the net electrostatic potential displays the
effects of the surface polarization of the droplet as it makes a
transition to the very different value that characterizes the whole
droplet.  It is clear in this case that the Ewald potential faithfully
captures this interior potential while avoiding detailed
considerations of the droplet interface.

{\bf Single ion hydration free energies are well-defined within
  molecular simulations}.  Electrostatic potentials can be defined
unambiguously as solutions of the Poisson equation with specified
charge densities and boundary conditions.  Electrostatic potentials
are computed throughout simulations of aqueous solutions even if
charge densities and boundary conditions may not be specified
explicitly.  If the system of interest is non-neutral, these issues
deserve emphasis because single ion free energies are typically not
measured experimentally.

On a molecular scale, dependences on specifics of the boundary
conditions in the definition of single ion hydration free energies can
be avoided.  This is accomplished by spherically integrating the
electrostatic potential over the charge density around a charge site
up to a distance where that potential saturates.  Using Ewald
interactions corresponds to such a spherical
integration.\cite{deLeeuw:80:a,Nijboer:88} Figure~\ref{fig:cluster}
shows results for electrostatic potentials obtained using two
different choices for the boundary conditions -- a solute-water
cluster and a periodic system.  The observed agreement is a
non-trivial computer experimental observation.

\section{System Size Extrapolation}
Computer simulations are performed for a finite system of molecules.
In most applications, the properties of the thermodynamic-limit,
infinite system are sought.  In coulombic systems, pronounced
finite-size effects are ubiquitous due to the long range of the
interactions.  We can separate finite-size effects in coulombic
systems into two categories: (1) those caused purely by the long-range
electrostatics independent of the thermodynamic state; and (2) finite
size effects that depend on the thermodynamic state (temperature,
pressure, etc.).

The electrostatic finite-size effects in the Ewald treatment of
Coulomb interactions arise from self-interactions and interactions
with the neutralizing background essential for non-neutral systems
under periodic boundary conditions.  Electrostatic finite-size effects
can be treated exactly by including the second and third sum in the
electrostatic energy $U_{\rm el}$ of the solute eq~\ref{eq:Uew_sol},
which account for self-interactions of the
solute.\cite{Hummer:95:e,Hummer:96:a,Hummer:93,Figueirido:95} For an
ion of charge $q$, the resulting correction to the solvation chemical
potential is\cite{Hummer:96:a,Hummer:93}
\begin{eqnarray}
  \label{eq:corr_elec}
  \mu_{\rm elec} & = & \mu_{\rm sim} + \frac{q^2 \xi}{2}~,
\end{eqnarray}
where $\mu_{\rm sim}$ is the chemical potential for charging the ion
from zero charge to net charge $q$ calculated from the Ewald
interactions with the solvent excluding self-interactions ({\em i.e.},
including only the first sum in eq~\ref{eq:Uew_sol}); $\mu_{\rm
  elec}$ includes the self-interactions; and $\xi$ is the ionic
self-term.  For a cubic box of length $L$, we have $\xi =
\lim_{r\rightarrow 0}[\varphi({\bf r})-1/|{\bf r}| \approx
-2.837297/L$.  Electrostatic finite size corrections for polar
molecules are developed in Ref.~\onlinecite{Hummer:95:e}.
The corresponding free energy of changing partial charges located at
positions ${\bf r}_{\alpha}$ on a molecule from $q_{\alpha}$ to
$q_{\alpha}'$ is
\begin{eqnarray}
  \label{eq:DFelec}
  \Delta \mu_{\rm elec} & = & \Delta \mu_{\rm sim} + \left\langle
    \frac{1}{2}\sum_{\alpha,\beta \atop \alpha \neq \beta}
    \left(q_{\alpha}' q_{\beta}' -
      q_{\alpha} q_{\beta}\right) \left[
      \varphi({\bf r}_{\alpha\beta})-\frac{1}{|{\bf
          r}_{\alpha\beta}|}\right]\right\rangle\nonumber\\\
  &&+\frac{1}{2}\sum_\alpha \left(q_{\alpha}'^2 -
      q_{\alpha}^2\right)\xi
\end{eqnarray}
where $\langle\ldots\rangle$ denotes a canonical average.  $\Delta
\mu_{\rm sim}$ includes only the energy difference corresponding to the
first sum in eq~\ref{eq:Uew_sol}, excluding self-interactions.  Note
that in constant pressure simulation with varying box volume, $\xi$
must also be averaged.

Thermodynamic finite-size effects on the other hand can only be
corrected approximately within a model.  For instance, we can use the
difference between a truly infinite version of a model and its finite
periodic version to correct for thermodynamic finite size
effects,\cite{Hummer:JCP:97,Figueirido:97,Lynden-Bell:97} as
schematically shown in
Figure~\ref{fig:fin_siz_scheme}.\nocite{Hummer:97} For spherical ions,
a Born model\cite{Born:20} and its periodic equivalent leads to finite
size corrections that depend on the dielectric constant $\epsilon$ of
the solvent, an effective Born radius $R_{\rm B}$ of the ion, and the
net charge $q$ of the ion:\cite{Hummer:JCP:97,BORN}
\begin{eqnarray}
  \label{eq:Born}
   \mu_{\rm therm} & \approx & \mu_{\rm elec} + \frac{1}{2} q^2
   \left[
     \frac{-\xi}{\epsilon} +
     \frac{4\pi (\epsilon-1) {R_{\rm B}}^2}{3\epsilon L^3}\right]~,
\end{eqnarray}
where $\mu_{\rm therm}$ is the chemical potential for charging that
includes the thermodynamic and electrostatic finite-size corrections.

Figure~\ref{fig:df} illustrates the finite size effects for the free
energies of charging Im(p) and Im(+) after correction for Ewald
self-interactions.  Free energies were calculated from sixth-order
integration formulas with corrected means and variances from
Table~\ref{tab:im_ave_var} that include the
electrostatic\cite{Hummer:96:a} but not the thermodynamic finite-size
correction. The free energy is plotted as a function of $1/L^3$ where
$L$ is the box length.  We find that the free energy of charging the
polar Im(p) is independent of the system size within the statistical
errors of about 1 kJ/mol for $N=16$ to $N=512$ water molecules.
However, the free energy of charging the Im(+) cation shows a system
size dependence proportional to $1/L^3$, as would be expected from our
finite size analysis.\cite{Hummer:JCP:97} Rather than using a more
realistic shape of the molecule in the dielectric model of
ref~\onlinecite{Hummer:JCP:97}, we fit the observed free energies to
to the spherical Born model eq~\ref{eq:Born} with an effective radius
$R_{\rm B}=0.207$ nm and an infinite dielectric constant which
reproduces the data over the whole range of system sizes of $16\leq
N\leq 512$ water molecules.

To further illustrate the power of the finite-size corrections,
Figure~\ref{fig:collapse} shows the probability distributions of
electrostatic interaction energies $U_{\rm el}$ of the imidazolium
cation Im(+) with the $N$ water molecules eq~\ref{eq:Uew_sol}.  Also
shown in Figure~\ref{fig:collapse} are the corresponding gaussian
distributions, which nicely reproduce the calculated histograms of
$U_{\rm el}$.  However, we observe a strong system size dependence:
small systems have narrower distributions of $U_{\rm el}$ with less
negative averages compared to large systems.  When we apply the
electrostatic and thermodynamic finite size corrections for the mean
and average, the gaussian distributions ``collapse'' to a single
distribution corresponding to the limit of an infinite system size.
We note that the finite-size correction is large: for $N=16$ water
molecules, the average $U_{\rm el}$ changes by about $-370$ kJ/mol
(250 $k_{\rm B}T$).

\section{Perturbation Theory}
A fundamental view of the thermodynamics due to electrostatic
interactions may be obtained from consideration of the distribution
$p(u;\lambda=0)$ of electrostatic energies $u$ in the reference charge
state $\lambda=0$.  The part of the chemical potential due to
electrostatic interactions $\Delta\mu(\lambda)$, the thermodynamic
parameter sought is then expressed by the fundamental result
\begin{equation}
  \label{eq:mu_exp}
  e^{-\beta\Delta\mu(\lambda)} =
  \left\langle e^{-\beta\lambda u} \right\rangle_{\lambda=0} = \int du \;
  p(u;\lambda=0) e^{-\beta\lambda u}.
\end{equation}
Here, $\beta^{-1}= k_{\rm B}T$ is Boltzmann's constant times the
temperature and $\langle\ldots\rangle_{\lambda=0}$ denotes a thermal
average with the solute in reference state $\lambda=0$. This formula
requires the consideration of the electrostatic potential even though
the electrostatic potential of a phase is an operationally subtle
property.\cite{Hummer:97,Ashbaugh:97} Despite that subtlety, the
potential sought is conceptually well-defined as the solution of
Poisson's equation with specified charge density and boundary
conditions.\cite{Hummer:96:a,Hummer:93,Hummer:97,Figueirido:95,%
Hummer:JPC:98}

Direct use of eq \ref{eq:mu_exp} can present difficulties.  Though
$p(u)$ is often substantially gaussian, the fundamental formula
eq~\ref{eq:mu_exp} is sensitive to the tails of $p(u)$.  That limits
the applicability of eq~\ref{eq:mu_exp} for calculations of even small
changes in the charge state $\lambda$. In addition, the simple
estimator $\ln\langle e^{-\beta\lambda u}\rangle_{\lambda=0} \approx
\ln [M^{-1}\sum_{i=1}^M e^{-\beta\lambda u_i}]$ from $M$ energies
$u_i$ observed in a simulation is biased and large sample sizes $M$
are required for this bias to be negligible.\cite{Wood:91}

Perturbation or cumulant expansions provide a technique to analyze
these distributions.\cite{Pratt:94:a,Hummer:95:e,Hummer:96:a,Levy:91,%
Hwang:87,Zwanzig:54,Smith:94,Zhou:95,Hummer:96:c,Liu:96,Archontis:96} A
cumulant expansion\cite{Kubo:62} with respect to $\lambda$ of
eq~\ref{eq:mu_exp} provides
\begin{eqnarray}
\left\langle \exp\left( -\beta\lambda u \right)
\right\rangle_{\lambda=0} & = & \exp\left[ \sum_{n=0}^{\infty}
  (-\beta\lambda)^n
\frac{C_n}{n!} \right]~.
\label{eq:dFC}
\end{eqnarray}
This defines the cumulants $C_n$ of order $n=0,1,2$ as
\begin{mathletters}
\begin{eqnarray}
C_0 & = & 0\\
C_1 & = & \left\langle u \right\rangle_{\lambda=0}\\
C_2 & = & \left\langle
        \left( u - \left\langle u \right\rangle_{\lambda=0} \right)^2
      \right\rangle_{\lambda=0}~.
\end{eqnarray} \end{mathletters}
We interpret eq~\ref{eq:dFC} as a Taylor expansion in $\lambda$ but
augment $\Delta\mu(\lambda)$ to include the self-contribution
$(\lambda q)^2 \xi / 2 $, where $\xi$ vanishes in the thermodynamic
limit but accounts for finite-size effects as discussed above.  Then
for the charging of an ion from a neutral reference condition we
have
\begin{eqnarray}
\Delta\mu(\lambda) & = & \lambda q  \left\langle u
\right\rangle_{\lambda=0} - \frac{(\lambda q)^2}{2}
\left[ \beta \left\langle \left(u - \left\langle u \right\rangle_{\lambda=0}
\right)^2 \right\rangle_{\lambda=0} -\xi \right]
+\cdots~.
\label{eq:dmuC}
\end{eqnarray}
This result should be compared to the Born\cite{Born:20} formula for
the hydration free energy due to electrostatic interactions of a
spherical ion of radius $R$ and charge $\lambda q$:
\begin{eqnarray}
\Delta\mu_{\rm B}(\lambda) & = & - {(\lambda q)^2 \over 2R} \left({\epsilon -1
\over \epsilon}\right)~,
\label{eq:born}
\end{eqnarray}
where $\epsilon$ is the dielectric constant of the solvent.  The
matching of the second-order terms between the cumulant expansion and
the continuum formula provides a determination of the radius R of a
spherical ion.  The distributions, $p(u;\lambda=0)$, required to evaluate the cumulant
averages involve the non-electrostatic interactions, $\lambda$=0, between solvent
molecules and the solute.  The indicated average thus generally
depends upon full characterization of the solvent.  Note that the
continuum model neglects the molecular contribution linear in
$\lambda$.  This linear term contributes to the asymmetry between
anion and cation solvation, making the solvation of anions more
favorable for a given ion size.\cite{Hummer:96:a}

In principle, higher-order cumulants could be used to obtain
information about the other Taylor coefficients.  However, as was
observed by Smith and van Gunsteren,\cite{Smith:94} higher-order
cumulants are increasingly difficult to extract from computer
simulations of limited duration.  Though direct extension of
perturbation theory beyond fourth order has been impractical,
interpolative approximations polynomial in $\lambda$ have been more
successful.  For the charging of water and ions, polynomials of order
six and higher were necessary to account for the simulation
data.\cite{Hummer:95:e,Hummer:96:a,Hummer:96:c} Thus, perturbation
theory was found to be unsatisfactory in such cases. For
atomic\cite{Hummer:96:a} and molecular ions,\cite{Garde:98}
a kink is typically observed for $d\Delta \mu(\lambda) /d \lambda$ as
a function of charge $\lambda$  at
modest values of this parameter when the solvation shell changes from
a cationic to an anionic structure.   Additional nonlinearities were
observed at high values of the ionic charge
$\lambda$.\cite{Jayaram:89,figueirido:94}

Table \ref{tab:im_ave_var} contains the averages and variances
corrected for electrostatic finite-size effects of the electrostatic
energies of Im(p) and Im(+) for system sizes between 16 and 512 water
molecules.  Errors of one standard deviation of the mean were
estimated by plotting the block error as a function of the number of
blocks.  The estimated error reaches a plateau when block values are
uncorrelated.  From the averages and variances, we can calculate the
chemical potentials of charging using integration formulas $(ijk)$
exact to various orders that involve $i$, $j$, and $k$ derivatives of
the free energy with respect to the coupling parameter at the
uncharged, half-charged, and fully charged state:\cite{Hummer:96:c}
\begin{mathletters}
  \begin{eqnarray}
    \label{eq:int}
    \Delta \mu(010) & \approx & C_1(\lambda=0.5) \label{eq:zoz}\\
    \Delta \mu(101) & \approx & \frac{1}{2}\left[
    C_1(\lambda=0)+C_1(\lambda=1)\right] \label{eq:ozo}\\
    \Delta \mu(111) & \approx & \frac{1}{6}\left[ C_1(\lambda=0) + 4
    C_1(\lambda=0.5) + C_1(\lambda=1)\right] \label{eq:ooo}\\
    \Delta \mu(202) & \approx & \Delta \mu(101) - \frac{\beta}{12}
    \left[C_2(\lambda=0)-C_2(\lambda=1)\right] \label{eq:tzt}\\
    \Delta \mu(212) & \approx & \frac{1}{30}\left[7 C_1(\lambda=0) + 16
      C_1(\lambda=0.5) + 7 C_1(\lambda=1)\right]\nonumber\\
    & & - \frac{\beta}{60} \left[C_2(\lambda=0)-C_2(\lambda=1)\right]
    \label{eq:tot}
  \end{eqnarray}
\end{mathletters}
where the cumulants $C_1$ and $C_2$ contain the electrostatic
finite-size corrections.  Formulas involving higher cumulants and
different nodes $\lambda_i$ are discussed in
Ref.~\onlinecite{Hummer:96:c}.  The integration formulas eqs
\ref{eq:zoz} to \ref{eq:tot} are exact to order 2, 2, 4, 4, and 6 in a
perturbation expansion, respectively.\cite{Zhou:95,Hummer:96:c}
Figure~\ref{fig:order} shows the free energy difference between Im(+)
and Im(p) as a function of the integration order for the 512 water
molecule system.  We find that as the order of the integration formula
increases, the free energy difference converges, with the sixth order
formula bracketed by the two fourth-order formulas.  The statistical
error of the free energy difference is about 1.5 kJ/mol.  Notice that
the discrepancy between the two second-order results of
Figure~\ref{fig:order} is significant on the scale of the statistical
uncertainties.  This emphasizes that the charging free energy is not a
quadratic function of the coupling parameter.  Note that the centered
second-order formula\cite{King:93} has a smaller systematic error.

Figure \ref{fig:TD_cycle} illustrates the complete four-node
thermodynamic cycle, where $\lambda$ is a coupling parameter changing
the partial charges on the molecule linearly from state zero to one.
The four nodes of the cycle are the uncharged and charged imidazole
and imidazolium.  We find that the free energies of charging and
conformational changes are consistent within the statistical errors.
Interestingly, the free energy of charging the polar Im(p) to the
Im(+) cation has a maximum for the linear charging path chosen here.
This increase reflects the linear terms of eq~\ref{eq:dmuC}, {\em i.e.},
increasing the net charge on the imidazole initially costs free
energy.

Dielectric continuum models predict a quadratic proportionality of the
free energy of charging on the linear coupling parameter $\lambda$.
In a molecular theory, such a quadratic charging free energy arises
when the probability density of electrostatic potential fluctuations
is gaussian.  Second-order perturbation theory would then be exact.
Figure~\ref{fig:taylor} compares second-order perturbation theory with
the reference sixth-order free energy polynomial calculated from the
averages and variances in Table~\ref{tab:im_ave_var}.  We find that
the perturbation expansions about the charged state ($\lambda=1$) are
accurate over a relatively wide range from $\lambda=1$ to almost
$\lambda=0.2$.  The expansion about the uncharged state $\lambda=0$ on
the other hand breaks down rapidly at about $\lambda=0.2$.

\section{Non-Gaussian Fluctuations}
{\bf Multistate Gaussian Models}. One idea for improvement of
dielectric models is based upon a physical description of the
structure of the first hydration shell.  It can be viewed from the
perspective of Stillinger-Weber inherent structures or
substates.\cite{Stillinger:82} These are potential energy basins of
attraction for steepest-descent quenching of first hydration shell
molecules.  If those first hydration shell molecules stayed always in
one basin, then a gaussian model for thermal fluctuations would be
reasonable.  Empirical radii parameters reflect the characteristics of
that single basin. However, changing conditions may result in
reweighting of slightly accessible basins or the opening of new
basins.  The gaussian or dielectric models may fail to describe these
possibilities well.  This picture is physically better defined than
the commonly nonspecific discussions of electrostriction and
dielectric saturation.

A corresponding ``multistate gaussian model'' was developed in
ref~\onlinecite{Hummer:97c}.  Attention is directed to the thermal
probability distribution of electrostatic potential energies of the
solute.  Rather than approximating this distribution as a single
gaussian distribution, perhaps with perturbative corrections, we
discriminate hydration structure on the basis of simple parameters
diagnostic of hydration substates.  We assume that the probability
distribution of electrostatic potential energies is gaussian for each
substate.  Therefore the full distribution is a superposition of
gaussian distributions for the various substates.

Thus we attempt to represent the observed complicating features of
$p(u)$ by a combination of simpler states:
\begin{eqnarray}
  \label{eq:expansion}
  p(u) & = & \sum_n w_n p_n(u)~,
\end{eqnarray}
with weights $w_n\geq 0, \sum_n w_n=1$ and normalized densities
$p_n(u)\geq 0, \int du\;p_n(u)=1$. We will seek $p_n(u)$'s of gaussian
form, representing the overall system as a linear combination of
gaussian subsystems, each showing linear response to electrostatic
interactions.  Representing $p(u)$ by a sum of gaussian densities can
give nontrivial results for the chemical potential, as can be seen by
substituting eq~\ref{eq:expansion} into eq~\ref{eq:mu_exp},
\begin{eqnarray}
  \label{eq:mu_gn}
  \Delta\mu(\lambda) & = & -k_B T
  \ln \sum_n w_n e^{-\beta\lambda
    m_n+\beta^2\lambda^2{\sigma_n}^2/2}~,
\end{eqnarray}
where $m_n$ and ${\sigma_n}^2$ are the mean and variance of the
gaussian $p_n$, respectively.

The non-gaussian fluctuations of the electrostatic potential in liquid
water are associated with changes in the conformations of protons that
make hydrogen bonds to the solute.  If those fluctuations could be
tempered, a gaussian model might become more accurate.  Thus, suitable
substate diagnostic parameters are the number of hydrogen bonds made
to the solute.

Explicit calculations have shown that this approach eliminates most of
the detailed numerical inaccuracies of the gaussian fluctuation models
for hydration of a water molecule in liquid water.\cite{Hummer:97c}
The markedly non-gaussian $p(u)$ was accurately represented as the sum
of gaussian distributions implied by this definition of a hydration substate.
We found $w_n>10^{-3}$ for $1\leq n\leq 6$ with 3.64 being the average
number of neighbors and $n=4$ the most probable number of neighbors.
The calculated change of the chemical potential upon change of the
charge state of a solute water molecule is correct to within 5 {\%}.
This is a remarkable result because $\Delta\mu(\lambda)$ is
non-quadratic, requiring an eighth-order polynomial to fit the
simulation data for chemical-potential
derivatives.\cite{Hummer:95:e,Hummer:96:a,Hummer:96:c} This shows that
sufficient information can be extracted from the simulation to
describe the distribution $p(u)$ helpfully; and that such an approach
can be successful even for  perturbations involving changes of
the chemical potential as large as 14~$k_{\rm B}T$.

Similar behavior can be anticipated for hydration of other neutral,
polar solutes such as the imidazole example studied here.
Figure~\ref{fig:taylor} (inset) shows the results of the multistate
gaussian model applied to charging and uncharging the polar imidazole
Im(p).  Fluctuation data were collected from a simulation of the
uncharged and charged Im(p) in $N=128$ water molecules, extending over
$10^6$ MC passes to allow for error estimates.  Instead of determining
the overall mean and variance of the electrostatic potential for a
second-order perturbation expansion, we calculate the means and
variances for several gaussian distributions from structures sorted
according to the number of hydrogen bonds.  Inspection of the radial
distribution functions of water oxygen and hydrogen around imidazole
sites shows one strong hydrogen bond donor, H1, and one acceptor, N3.
As a criterion for the formation of a water-imidazole hydrogen bond,
we used that the distance between the acceptor (nitrogen N3 or water
oxygen) and the donor hydrogen has to be smaller than 0.23 nm.  We can
then sort structures according to the numbers of hydrogen bonds
accepted and donated by the solute.  With this simple criterion, we
find that six and two gaussian distributions contribute to the
expansions about the charged and uncharged state of Im(p),
respectively.  This multistate gaussian model greatly improves the
quality of the expansions, both about the charged and uncharged state.
The reference free energy is now within the statistical errors of the
two multistate gaussian models over the whole range $0\leq\lambda\leq
1$.


{\bf Quasi-chemical theories}.  Those more difficult anionic cases
mentioned above can be attacked more directly.  The local neighborhood
is again used to discriminate structural possibilities.  But, in
addition, the consequences for the hydration free energy of the
molecular interactions within that neighborhood are treated fully.
This reserves the longer-ranged interactions for simple
approximations, {\it e.g.\/} with gaussian models.

These theoretical developments arose from recent molecular
calculations\cite{Martin:97} that suggested how a chemical perspective
can be helpful in computing thermodynamic properties of water and
aqueous solutions. That calculation used electronic structure results
on the Fe(H$_2$O)${_6}^{3+}$ cluster and simple, physical estimates
of further solvation effects.  The results were organized according to
the pattern of a simple chemical reaction and a surprisingly accurate
evaluation of the hydration free energy was obtained.  Despite this
recent motivation, the theories developed are akin to good
approximations of historical and pedagogical importance in the areas
of cooperative phenomena and phase transitions.\cite{brush} In those
areas, similar approximations are called Guggenheim, quasi-chemical,
or Bethe approximations.

These quasi-chemical theories\cite{pratt:98} are constructed by
considering a geometric volume fixed on the solute molecule and
performing a calculation analogous to the evaluation of a grand
canonical partition function for that volume.  All the possibilities
for occupancy of that volume by solvent or other solution species must
be considered eventually.  The final result can be described by
reference to simple patterns of chemical equilibria such as:
\begin{eqnarray}
X^{q-} + n H_2O \rightleftharpoons X(H_2O)_n{}^{q-}~.
\label{equilibrium}
\end{eqnarray}
The solute of interest is denoted here generically as $X^{q-}$; a
star-type cluster of that solute with $n$ water (W) molecules is
considered as the product.  For such a cluster, call it an M-cluster,
we could calculate the equilibrium ratio $K_M$ for a dilute gas phase.
Note that $K_M\rho_W {}^n$ is dimensionless and this observation
resolves standard state issues.  The factors denoted by $\langle
\exp\{-\beta \Delta u\}\rangle_{0,C}$, where $C$ indicates either a
water molecule or the M-cluster, carry information about solvation
free energy of the species involved.  For the species other than the
cluster this is the familiar Widom factor.  For the cluster, this
factor requires slight additional restriction but can be verbally
defined by saying that it is the average of the Boltzmann factor for
cluster-solution interactions over the thermal motion of the cluster
and solution under the condition that the only interactions between
these subsystems rigidly exclude additional solvent molecules from the
cluster volume for the complex.  This restriction enforces a
constraint required to preserve simple enumerations that underlie
these results.  The theoretical structure is designed so that simple
approximations such as dielectric models might be used for the factors
$\langle
\exp\{-\beta \Delta u\}\rangle_{0,C}$.  But more detailed techniques
might be applied to the calculation of $K_M$.  Finally, we compile
\begin{eqnarray}
{\tilde K}_M \equiv K_M { \langle \exp\{-\beta \Delta
u\}\rangle_{0,M} \over [ \langle \exp\{-\beta \Delta u\}\rangle_{0,W}
]^n }~.
\label{k-tilde}
\end{eqnarray}
Thus, this ${\tilde K}_M$ is built on the pattern of the chemical
equilibrium eq \ref{equilibrium} but without a `solvation factor' for
the `bare' solute.

Now consider all possibilities for clusters.  A thermodynamic
implication of this information is:
\begin{eqnarray} \mu_{X^{q-}} =  kT\ln [{ \rho_{X^{q-}} V/ q_{X^{q-}}}
] - kT \ln [ p_0 \sum_{M} {\tilde K}_M \rho_W{}^n ] .
\label{answer6}
\end{eqnarray}  $p_0$ is the probability that the clustering volume
would be observed to be {\em empty} in the equilibrium solution; thus
$-kT\ln p_0$ is the free energy for formation of a cavity for the
clustering volume in the solution.  The sum is over all clusters with
zero or more ligands.  The product of the densities involved with each
term includes a density factor for each ligand.  This formula makes
the conventional separation between the contributions of
intermolecular interactions and the non-interaction (ideal) terms;
$q_{X^{q-}}$ is the partition function of the bare solute in the
absence of interactions with any other species and $\rho_{X^{q-}}$ is
the density of the solute.  As an example, for an atomic ion such as
the chloride ion Cl$^-$ we would put $q_{X^{q-}}=V/\Lambda^3$ with $V$
the volume of the system and $\Lambda$ the thermal deBroglie
wavelength of the chloride ion.  This formula becomes approximate when
approximate models are adopted for $p_0$, for $K_M$, and for the
solvation factors.  Those quantities depend on definition of the
clustering volume.  But, since the physical problem is independent of
those parameters, the theory should be insensitive to them.

The motivation of this approach is the fact that a substantial but
intricate part of the free energy sought is to be found in $K_M$.
The number of possibilities for ligand populations will be small for
molecular scale clustering volumes.  So a limited number of terms must
be considered.  Because the clusters will be small systems, elaborate
computational methods can be applied to the prediction of the
$K_M$, including current electronic structure techniques.  With the
complicated chemical interactions separated for individual treatment
the remaining hydration contributions should be simpler and the
required theories better controlled.

Equation \ref{answer6} should be compared with eq~\ref{eq:mu_gn}.  One
difference is that eq \ref{answer6} attempts to provide the whole
hydration free energy, not just the part due to electrostatic
interactions.  That explains the presence of $p_0$ in eq
\ref{answer6}.  Beyond that, the structures of these formulas are
similar.  The presence of more than one term in the sum of eq
\ref{answer6} is an expression of an entropy
contribution associated with the possibilities for different ligand
populations.  Finally, the complete calculation of the $K_M$
includes non-gaussian statistical possibilities not anticipated by
eq~\ref{eq:mu_gn}.

\section{Conclusions}
Recent calculations of the hydration free energy due to electrostatic
interactions between charged and polar solutes in water have obtained
high accuracy results for the simple molecular models that are the
basis of most simulation
calculations.\cite{Hummer:95:e,Hummer:96:a,Hummer:97c,Lynden-Bell:97,%
Kalko:96,Sakane:98} An important step in securing those high accuracy
results has been a careful consideration of treatment of long-ranged
interactions.  That work suggests that {\em the Ewald method is an
easy way to get correct hydration free energies} from molecular
calculations, that is, to achieve well characterized results
appropriate to the thermodynamic limit in which the system size tends
to infinity for given densities and temperature.  Additionally, it
suggests that molecular simulations with Ewald potentials and periodic
boundary conditions can have efficiencies comparable to rougher
approximations that are often employed to compute hydration free
energies for molecularly well-defined problems.\cite{CPU} And
furthermore, this has produced a simple, effective, and clear
understanding of how to extrapolate electrostatic hydration free
energies to the thermodynamic limit; an accurate evaluation of the
hydration free energies of imidazole and imidazolium can be obtained
with as few as 16 water molecules included in the simulation.

These high accuracy results for well-defined models permitted careful
testing of simple theories of electrostatic hydration free energies.
The simplest theories, dielectric continuum models, have been found to
be rough despite the fact that they can always be adjusted to
reproduce an empirical answer given {\it a priori.} Such a conclusion
has surely been widely expected.  However, the testing has led to new
theories, the multistate gaussian and quasi-chemical theories, that
should permit more revealing molecular scale calculations.  The
quasi-chemical approaches seem to provide the most natural way to
utilize current electronic structure packages to study electronic
structure issues for solution species.  This should be particularly
helpful for treatment of basic, molecular anions that are ubiquitous
in aqueous solution chemistry.

Physical conclusions more specifically are that the most prominent
failings of the simplest theories are associated with solvent proton
conformations that lead to non-gaussian fluctuations of electrostatic
potentials.  Thus, the most favorable cases for the second-order
perturbation theories are monoatomic positive ions.  In such cases,
oxygen-hydrogen bonds are oriented away from the ion, placing those
protons as far out as possible.  Neutral, polar molecules that may
form specific hydrogen bonds with the solvent are more difficult for
these theories, though the hydration free energies sought are smaller
in magnitude.  Negative molecular ions are expected to offer further
complications because now the problematic proton motions occur close
to the solute and the hydration effects will be larger for anionic
species.

\section*{Acknowledgment}  This work was supported by the LDRD
program at Los Alamos.  We thank G. J. Tawa for providing unpublished
results of his calculations on imidazole and imidazolium.
G.H. wants to thank Dr. Attila Szabo for many useful discussions and
collaborations.


\begin{thebibliography}{10}
\bibitem{BORN} Note the factor $4\pi$ in eq~\protect\ref{eq:Born},
  corrected from $2\pi$ in the original
  publication.\protect\cite{Hummer:JCP:97} A consequence of this
  correction is that, for sodium
  ions,\protect\cite{Hummer:JCP:97} the ion radius entering
  Born's free energy formula and the effective radius $R_{\rm B}$
  entering the correction formula eq~\protect\ref{eq:Born} are not
  identical.  Accordingly, the solute-size dependent term in the
  thermodynamic finite-size correction eq~\protect\ref{eq:Born} is
  best seen as giving the dimensional behavior proportional to
  $1/L^3$, with the solute size $R_{\rm B}$ as a parameter.
\bibitem{CPU}
The total CPU time of all imidazole and imidazolium simulations with
$N=16$ and 32 water molecules was, including equilibrations, 4 and 8
hours on a single SGI R10000 chip, respectively. Extrapolating the
$N=16$ and 32 results to $N\rightarrow\infty$, we obtain an effective
Born radius of 0.30 nm and a solvation free energy difference between
Im(+) and Im(p) that is within 1.2 kJ/mol of the fit to all data for
$16\leq N\leq 512$.
\end{thebibliography}

\begin{figure}[htbp]
  \caption{Protonation equilibrium between imidazole and imidazolium.}
  \label{fig:imidazole-scheme}
\end{figure}

\begin{figure}
  \caption{Comparison of dielectric models (ordinate) with molecular
simulations (abscissa) for the induced electrostatic potentials due to
the solvent at the atom centers for Im(+) (upper panel) and Im(p)
(lower panel).  Dielectric model results were obtained for several
sets of radii in current use: diamonds: R$_C$=0.267~nm,
R$_N$=0.231~nm;\protect\cite{Nina} large circles: R$_{H(N)}$=0.1160
nm, R$_{H(C)}$=0.1710 nm, R$_C$=0.230~nm,
R$_N$=0.150~nm;\protect\cite{topol} small circles: R$_H$=0.1172~nm,
R$_C$=0.2096~nm, R$_N$=0.1738~nm;\protect\cite{Truong} crosses:
R$_H$=0.1172~nm, R$_C$=0.1635~nm,
R$_N$=0.1738~nm.\protect\cite{Truong}  A boundary element method was
used for the dielectric model calculations.\protect\cite{Pratt:97}
Notice that a radii set that happens to be qualitatively satisfactory
for the cation (diamonds) can be significantly less satisfactory for
the slightly different circumstance of the neutral polar molecule.}
\label{fig:compare}
\end{figure}

\begin{figure}[htbp]
  \caption{Electrostatic potential at the center of a neutral
    Lennard-Jones solute in SPC water\protect\cite{Berendsen:81} from
    simulations of a solute at the center of a cluster with 1024 water
    molecules (dashed line) and with periodic boundary conditions and
    Ewald summation (solid line).  Shown is the potential obtained by
    integrating the charge density around the solute up to a distance
    $R$ using $1/r$ (cluster) and $\varphi({\bf r})$ (Ewald) for the
    Coulomb interactions.  The results are those of
    ref~\protect\onlinecite{Hummer:97}.}
  \label{fig:cluster}
\end{figure}

\begin{figure}[htbp]
  \caption{Schematic representation of the thermodynamic finite-size
    correction.  The thermodynamic finite-size correction $\{\ldots\}$
    is the difference between an infinite Born model and a Born model
    under periodic boundary conditions.  A spherical ion of charge $q$
    and radius $R_{\rm B}$ is embedded in a medium with a dielectric
    constant $\epsilon$ inside the simulation box.  In addition, the
    box is filled with the neutralizing background charge.  Periodic
    boundary conditions are applied.  The corresponding electrostatic
    potential is determined from the Poisson equation with appropriate
    boundary conditions on the box boundary and ion
    surface.\protect\cite{Hummer:97}}
  \label{fig:fin_siz_scheme}
\end{figure}

\begin{figure}[htbp]
  \caption{Finite-size dependence of the free energies of charging
    imidazole and imidazolium (filled circles and open squares on the
    right and left hand scale, respectively), as a function of the
    inverse volume of the simulation box, $1/L^3$. The top scale gives
    the number of water molecules.}
  \label{fig:df}
\end{figure}

\begin{figure}[htbp]
  \caption{Finite-size correction of the probability
    densities $p(U_{\rm el})$ of the electrostatic energies $U_{\rm
      el}$ of Im(+).  The uncorrected $U_{\rm el}$ histograms are
    shown with symbols, together with corresponding gaussian
    distributions.  After correction for electrostatic and
    thermodynamic finite-size effects, the corresponding gaussian
    distributions ``collapse'' and agree closely for all system sizes
    of $16\leq N\leq 512$ water molecules.}
  \label{fig:collapse}
\end{figure}

\begin{figure}
  \caption{Free energy of charging the polar imidazole Im(p) to the
    imidazolium cation Im(+) as a function of the order of the
    integration formula.  $(ijk)$ indicates the number $i$, $j$, and
    $k$ of derivatives used at the uncharged, half-charged, and fully
    charged state.\protect\cite{Hummer:96:c}}
\label{fig:order}
\end{figure}

\begin{figure}[htbp]
  \caption{Thermodynamic cycle with the free energies connecting the
    four states of the uncharged and charged imidazole and imidazolium
    as a function of a linear coupling parameter. Shown are results
    for charging of imidazolium (solid line), uncharging of imidazole
    (long dashed line), conversion of imidazole to imidazolium (dotted
    line) and conversion of uncharged imidazole to uncharged
    imidazolium (short dashed line).}
  \label{fig:TD_cycle}
\end{figure}

\begin{figure}[htbp]

  \caption{Comparison of the second-order perturbation expansion
(dashed lines) with the reference free energies of charging Im(p) to
Im(+) (top panel), uncharged Im(+) to the cationic Im(+) (middle
panel), and uncharged Im(p) to the polar Im(p) (bottom panel).  Also
included as an inset in the bottom panel is a comparison with
multistate gaussian models (symbols and dot dashed lines) shown with
estimated statistical errors.  The multistate expansions about the
charged and uncharged states are shown with open squares and filled
circles, respectively.} \label{fig:taylor}
\end{figure}

\begin{table}[htbp]
  \begin{tabular}{rrrrrr}
    atom & $x$ & $y$ & $q$ & $\sigma$ & $\epsilon$ \\\hline
    \multicolumn{6}{c}{Imidazole}\\\hline
    N1 & $ 0.0000$ & $ 0.1105$ & $-0.090285$ & $0.325000$ & $0.71128$\\
    C2 & $-0.1091$ & $ 0.0282$ & $ 0.232373$ & $0.339967$ & $0.35982$\\
    N3 & $-0.0741$ & $-0.0983$ & $-0.715903$ & $0.325000$ & $0.71128$\\
    C4 & $ 0.0636$ & $-0.0984$ & $ 0.217356$ & $0.339967$ & $0.35982$\\
    C5 & $ 0.1120$ & $ 0.0298$ & $-0.374687$ & $0.339967$ & $0.35982$\\
    H1 & $-0.0009$ & $ 0.2112$ & $ 0.318027$ & $0.106908$ & $0.06569$\\
    H2 & $-0.2102$ & $ 0.0661$ & $ 0.102391$ & $0.242146$ & $0.06276$\\
    H4 & $ 0.1197$ & $-0.1905$ & $ 0.082346$ & $0.242146$ & $0.06276$\\
    H5 & $ 0.2119$ & $ 0.0700$ & $ 0.228383$ & $0.242146$ & $0.06276$\\\hline
    \multicolumn{6}{c}{Imidazolium}\\\hline
    N1 & $ 0.0000$ & $ 0.1128$ & $-0.115106$ & $0.325000$ & $0.71128$\\
    C2 & $-0.1086$ & $ 0.0353$ & $ 0.010825$ & $0.339967$ & $0.35982$\\
    N3 & $-0.0663$ & $-0.0912$ & $-0.122786$ & $0.325000$ & $0.71128$\\
    C4 & $ 0.0719$ & $-0.0949$ & $-0.139642$ & $0.339967$ & $0.35982$\\
    C5 & $ 0.1140$ & $ 0.0344$ & $-0.122097$ & $0.339967$ & $0.35982$\\
    H1 & $-0.0018$ & $ 0.2141$ & $ 0.398875$ & $0.106908$ & $0.06569$\\
    H2 & $-0.2110$ & $ 0.0686$ & $ 0.230198$ & $0.242146$ & $0.06276$\\
    H3 & $-0.1274$ & $-0.1721$ & $ 0.402905$ & $0.106908$ & $0.06569$\\
    H4 & $ 0.1273$ & $-0.1872$ & $ 0.232002$ & $0.242146$ & $0.06276$\\
    H5 & $ 0.2131$ & $ 0.0766$ & $ 0.224826$ & $0.242146$ & $0.06276$\\
  \end{tabular}
  \caption{Coordinates $x$ and $y$ (in nm) and charges $q$ (in
    elementary charge units $e$) of
    the atoms in the planar imidazole and imidazolium from to the quantum
    mechanical calculations of Topol et al.\protect\cite{topol}  The
    Lennard-Jones parameters $\sigma$ and $\epsilon$ (in nm and
    kJ/mol) are taken from the AMBER force field.\protect\cite{Cornell:JACS:95}
    Lorentz-Berthelot mixing rules were
    applied to combine the Lennard-Jones parameters of the solute
    atoms with those of SPC/E water.\protect\cite{Berendsen:87}}
  \label{tab:imidazole}
\end{table}

\begin{table}[htbp]
  \begin{tabular}{ccccccc}
     & \multicolumn{2}{c}{uncharged} & \multicolumn{2}{c}{half-charged}
    & \multicolumn{2}{c}{charged} \\
$N$ & ave & var & ave & var & ave & var\\\hline
\multicolumn{7}{c}{Imidazole}\\\hline
 16 & $   1.2\pm 0.5$ & $209.6\pm  7$ & $ -56.7\pm 1.3$ & $436.8\pm 25$ & $-156.9\pm 1.5$ & $495.6\pm 32$\\
 32 & $   0.5\pm 1.0$ & $224.4\pm  7$ & $ -57.\pm 1.6$ & $427.4\pm 22$ & $-158.7\pm 1.5$ & $456.4\pm 26$\\
 64 & $   1.8\pm 0.6$ & $208.2\pm  7$ & $ -58.8\pm 2.0$ & $510.0\pm 25$ & $-155.5\pm 1.5$ & $480.9\pm 15$\\
128 & $   1.0\pm 0.7$ & $205.1\pm  9$ & $ -59.2\pm 1.6$ & $429.5\pm 16$ & $-156.1\pm 1.8$ & $491.9\pm 20$\\
256 & $   1.3\pm 0.5$ & $204.8\pm  7$ & $ -59.2\pm 1.6$ & $429.1\pm 22$ & $-152.9\pm 2.0$ & $473.4\pm 24$\\
512 & $   1.1\pm 0.5$ & $209.6\pm  9$ & $ -57.7\pm 1.3$ & $410.8\pm
 17$ & $-155.9\pm 2.0$ & $539.2\pm 30$\\\hline
\multicolumn{7}{c}{Imidazolium}\\\hline
 16 & $  33.4\pm 1.3$ & $1326.7\pm 10$ & $-226.5\pm 1.0$ & $1305.5\pm  7$ & $-500.8\pm 1.0$ & $1391.0\pm 14$\\
 32 & $  33.9\pm 1.0$ & $1267.6\pm 12$ & $-215.0\pm 0.9$ & $1267.0\pm 10$ & $-483.4\pm 2.0$ & $1375.0\pm 20$\\
 64 & $  35.9\pm 1.1$ & $1234.0\pm 12$ & $-208.7\pm 1.3$ & $1252.4\pm 13$ & $-473.9\pm 1.5$ & $1307.9\pm 18$\\
128 & $  34.9\pm 1.0$ & $1226.0\pm 14$ & $-203.5\pm 0.9$ & $1215.9\pm 17$ & $-469.1\pm 1.3$ & $1291.2\pm 22$\\
256 & $  34.1\pm 1.1$ & $1222.5\pm 22$ & $-204.4\pm 1.2$ & $1256.3\pm 24$ & $-465.2\pm 1.5$ & $1330.2\pm 33$\\
512 & $  35.9\pm 1.5$ & $1248.8\pm 18$ & $-201.2\pm 1.8$ & $1248.3\pm 26$ & $-461.7\pm 1.5$ & $1305.3\pm 25$\\
  \end{tabular}
  \caption{Averages $C_1$ (in kJ/mol) and variances $C_2$ [in
    (kJ/mol)$^2$] of the
    electrostatic energy of imidazole and imidazolium in the
    uncharged, half-charged, and fully charged states.  Finite-size
    corrections have been applied.}
  \label{tab:im_ave_var}
\end{table}

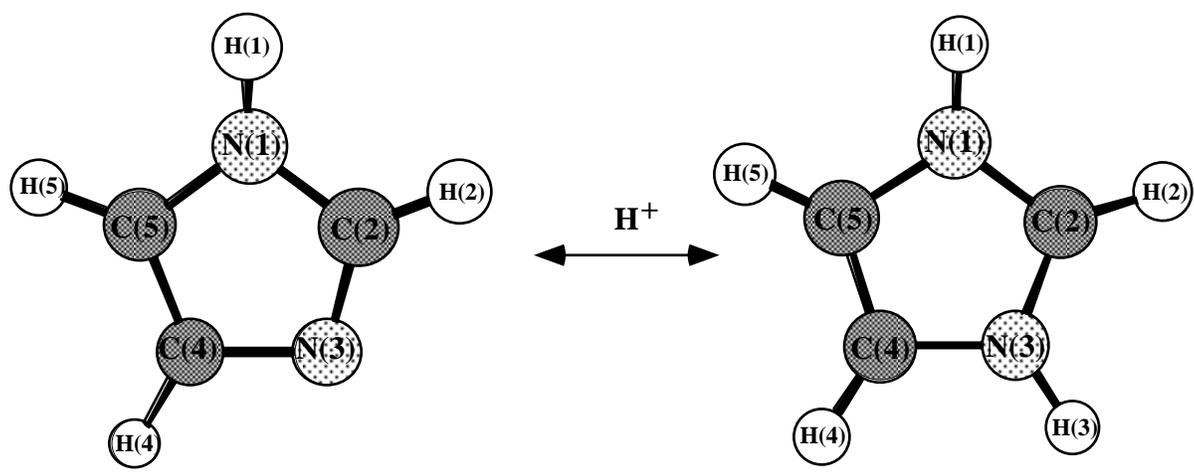
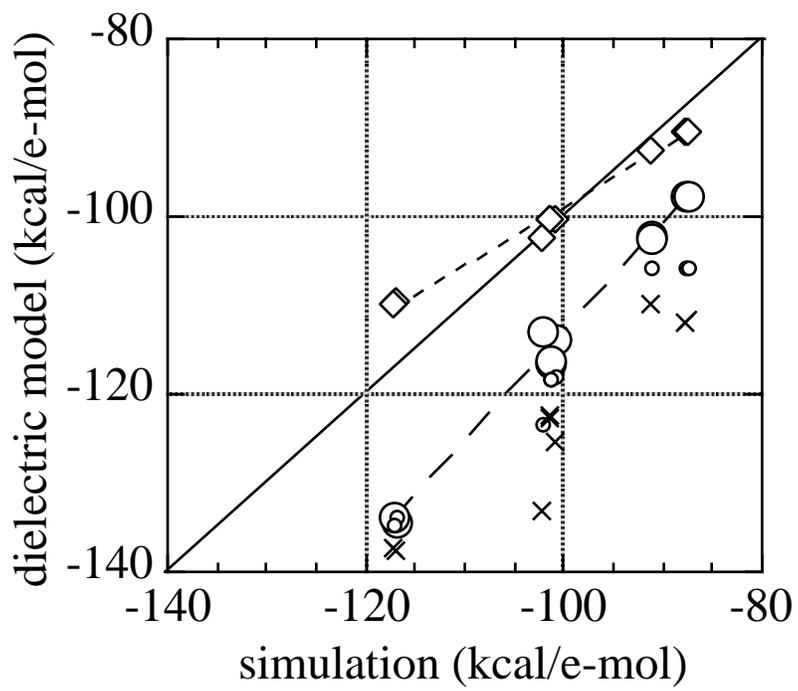
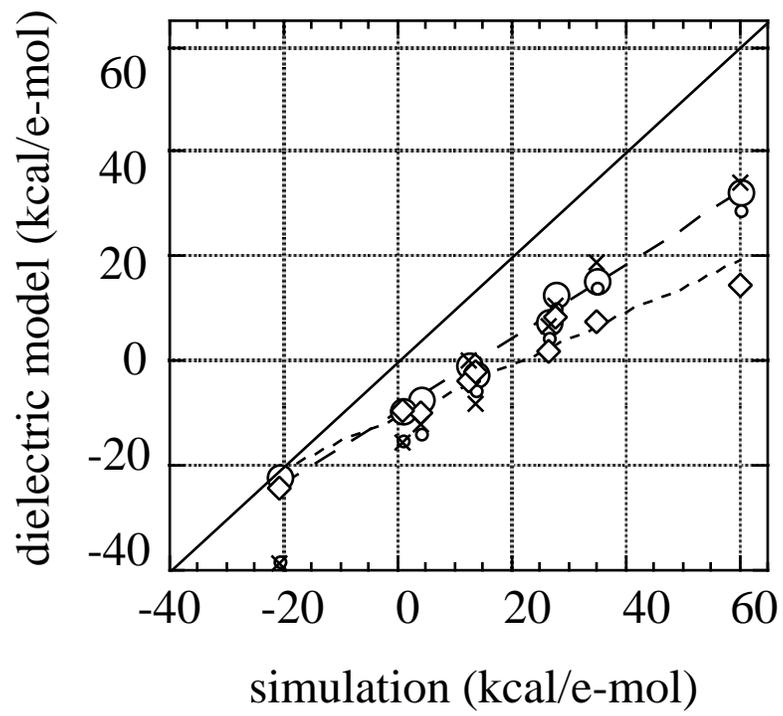
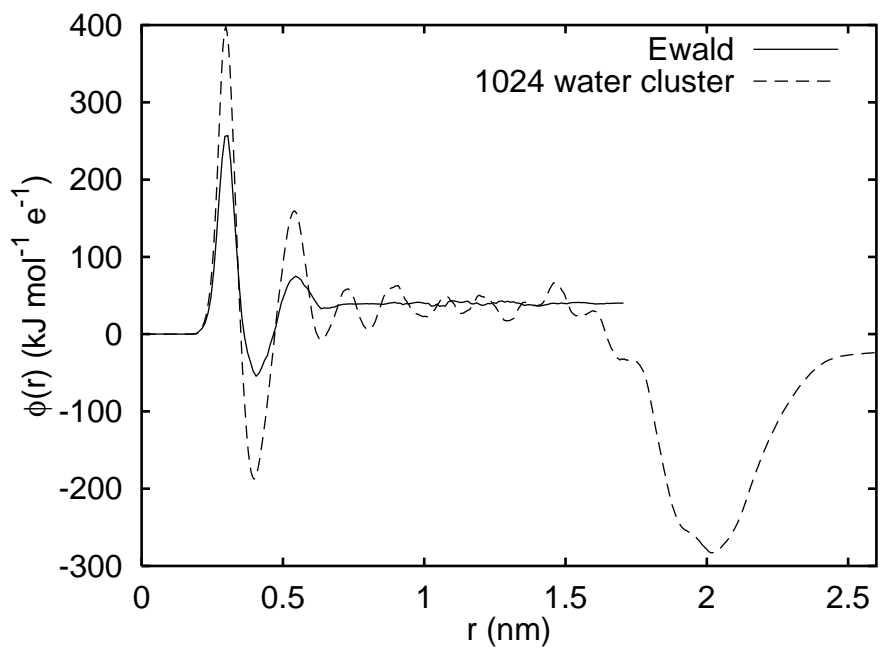
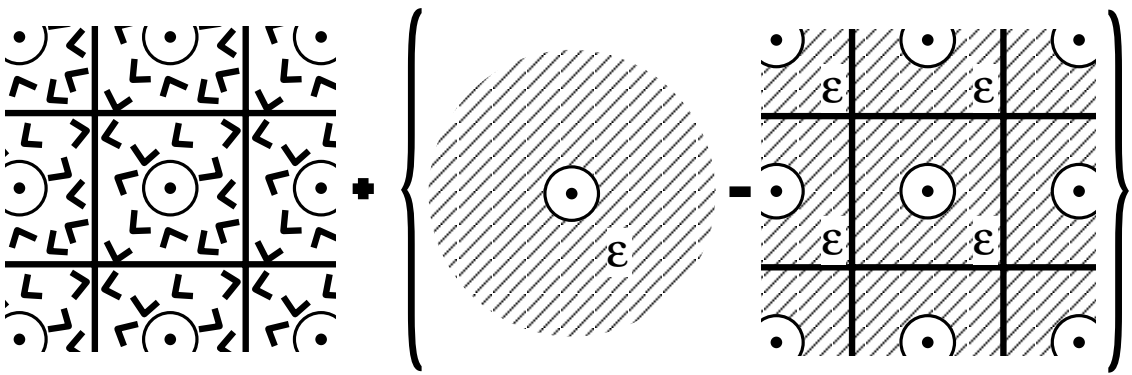
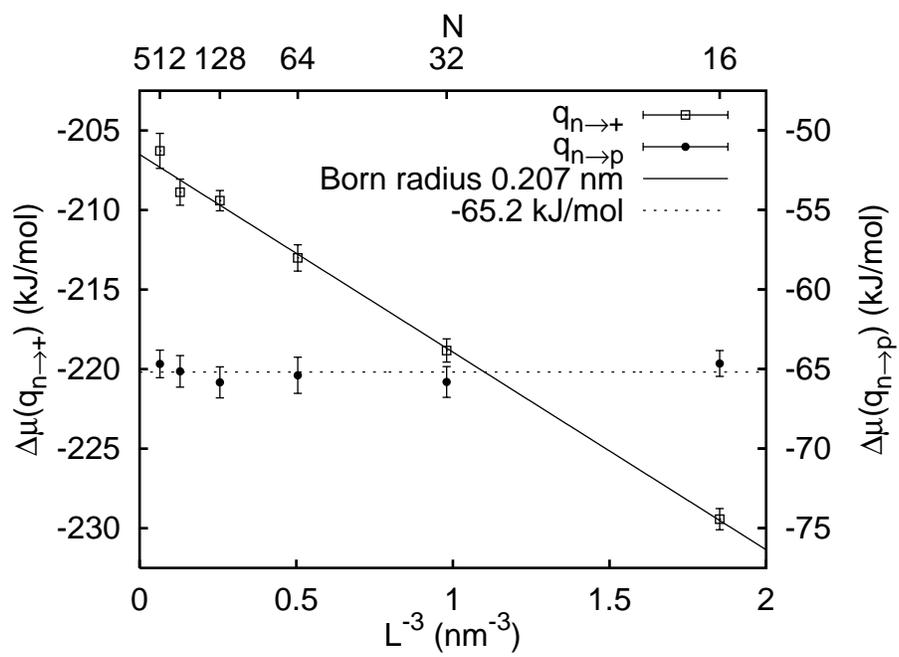
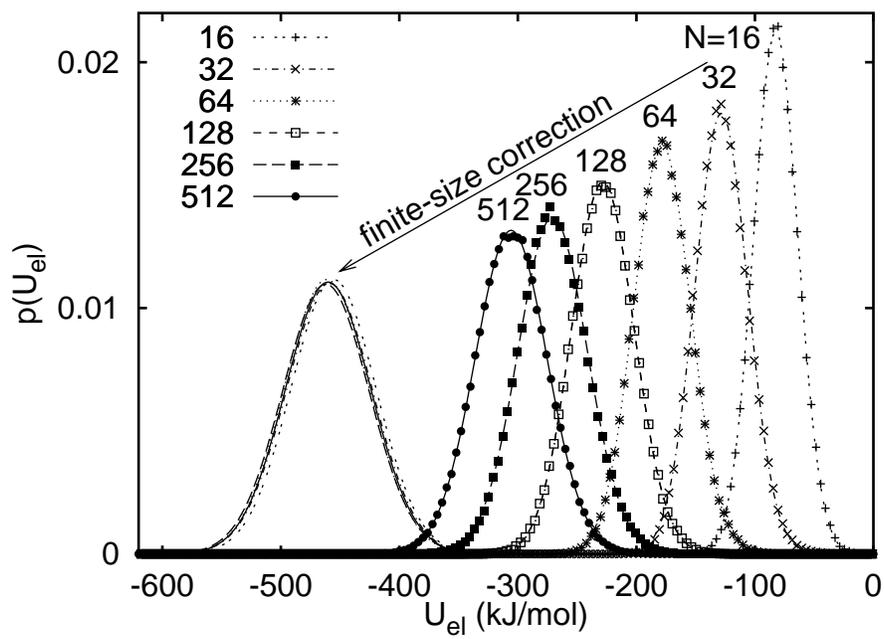
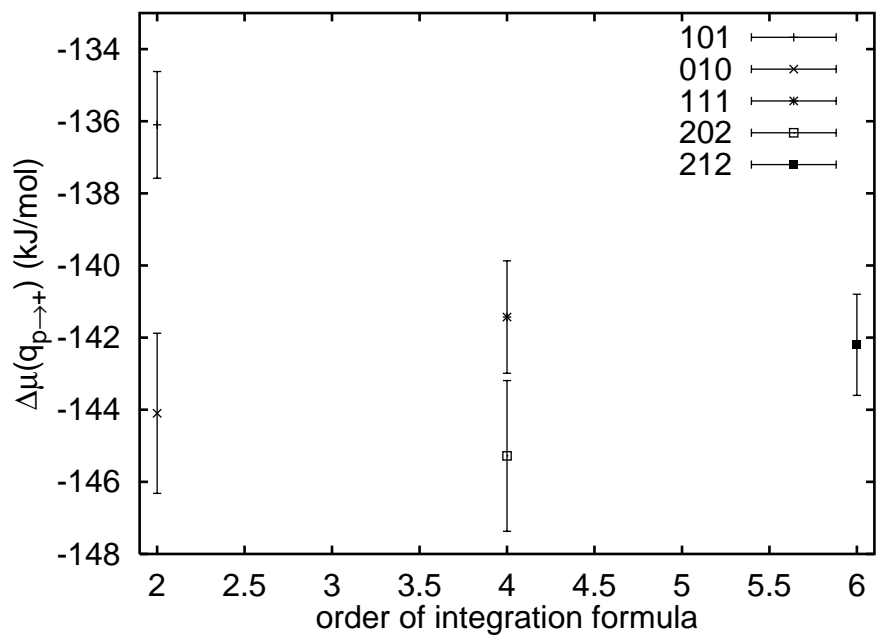
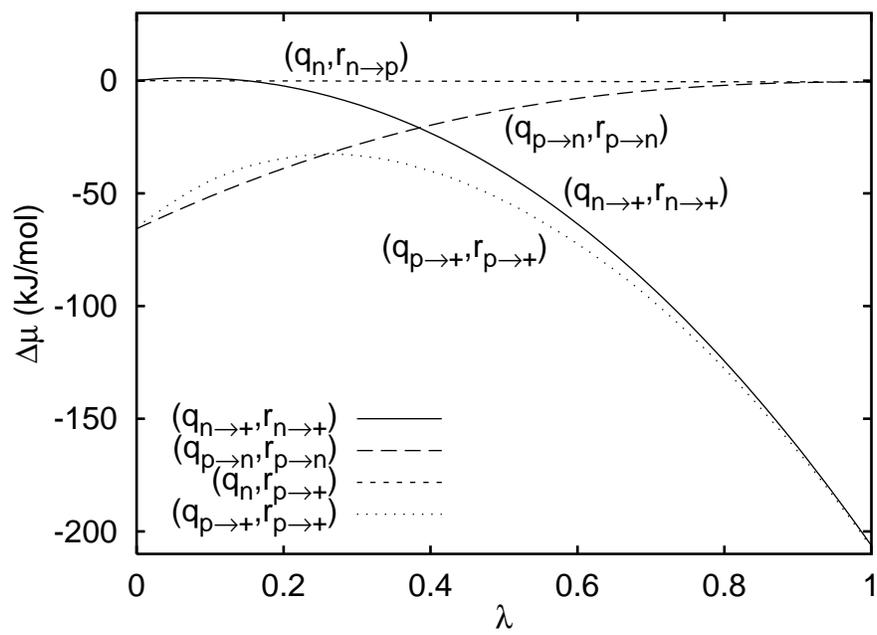
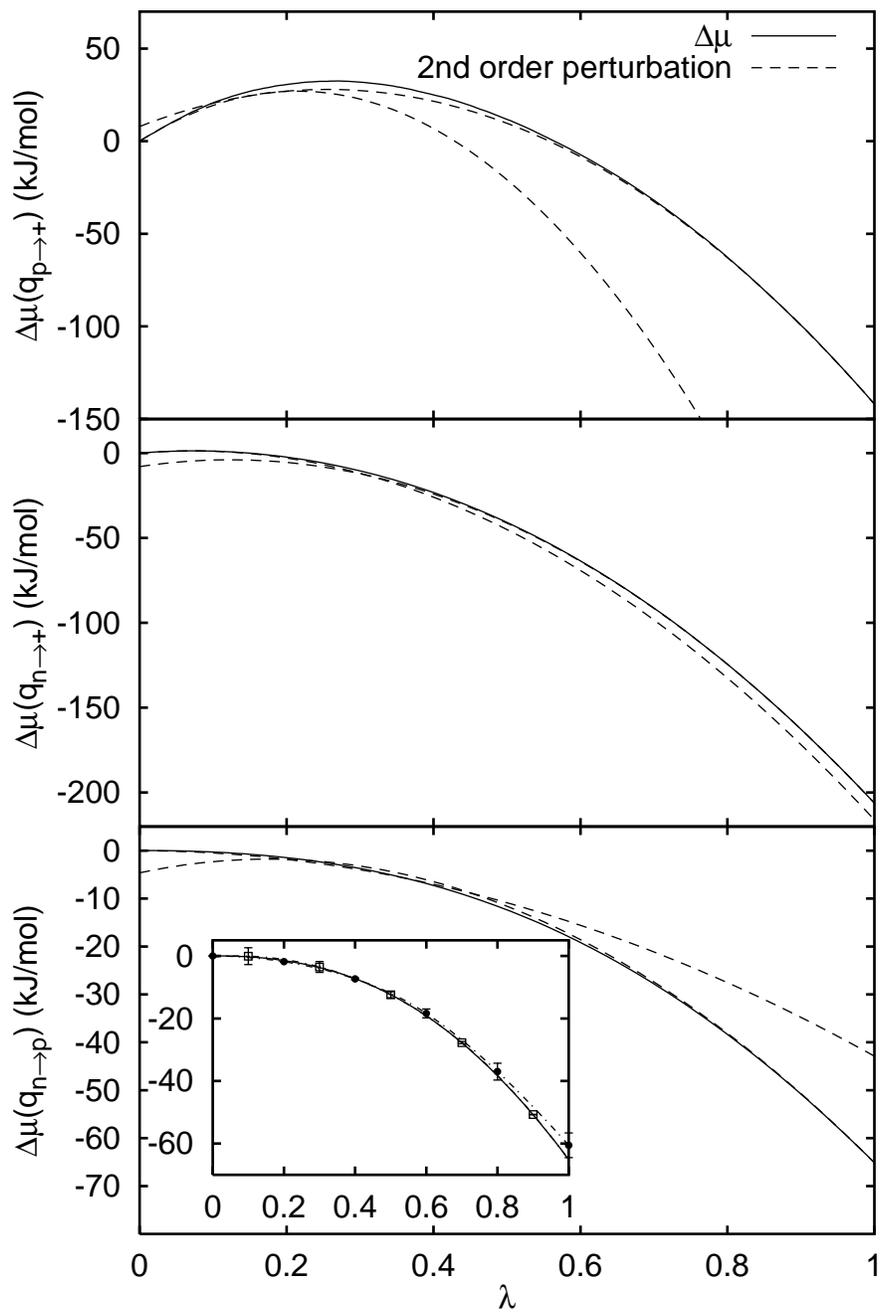
\end{document}